\documentclass{article}
\usepackage{epsfig}
\usepackage{multirow}
\usepackage{amsmath,mathtools}

\usepackage{amssymb}
\date{}
\def\d{\delta}
\def\f{\frac}
\def\app{\approx}
 
\def\p{\partial}

\begin{document}

\title{Exact theory and numeric results for short pulse ionization of simple model atom 
in one dimension}

\author{A.Rokhlenko 
\medskip
\\Department of Mathematics,
Rutgers University\\ Piscataway, NJ 08854-8019}

\maketitle
\begin{abstract}

Our exact theory for continuous harmonic perturbation of a one dimensional model atom by
parametric variations of its potential is generalized for the cases when a) the atom is 
exposed to short pulses of an external harmonic electric field and b) the forcing is 
represented by short bursts of different shape changing the strength of the binding potential. 
This work is motivated not only by the wide use of laser pulses for atomic ionization, but
also by our earlier study of the same model which successfully described the ionization 
dynamics in all orders, i.e. the multi-photon processes, though being treated by the 
non-relativistic Schr\"odinger equation. In particular, it was shown that the bound atom cannot 
survive the excitation of its potential caused by any non-zero frequency and amplitude of the 
continuous harmonic forcing. Our present analysis found important laws of the atomic ionization 
by short pulses, in particular the efficiency of ionizing this model system and presumably 
real ones as well.

\end{abstract}

We study a simple one-dimensional quantum system with the attractive potential,
modeled by the $\delta$-function. This system is assumed to be in the bound state until at
some initial time $t=0$ a) it becomes exposed to an external harmonic electric field or b) the  
strength of the system binding potential gets time dependent. The perturbation after a short 
interval $T$ is turned off and our objective is to study the time evolution of the bound state 
on the interval $0<t<T$. The pulses of external electric field are modeling the application of 
laser beams for atomic ionization. 

Excitation of the $\delta$-function atom was studied in [1-6] for a simpler case of harmonic 
parametric perturbation of its potential when $T=\infty$, i.e. of infinite duration, and the main 
conclusion was the complete ionization for arbitrary frequency and amplitude of perturbation. 
Other results have shown a surprising similarity of main features of the process with observed 
experimentally and numerically in spite of simplicity of this model.
\vskip0.3cm
\noindent
{\bf 1. PROBLEM SET UP}
\vskip0.2cm

As in [1] we start by considering the one-dimensional stationary system located at $x=0$ with 
an unperturbed Hamiltonian $$
H_0=-\f{\hbar^2}{2m}\f{d^2}{dx^2}-g\delta(x),\ g>0,\ \ -\infty <x< \infty, \eqno(1)$$
which has a single bound state$$
u_b(p,x)=\sqrt{p}e^{-p|x|},\ \ p=\f{m}{\hbar^2}g. \eqno(2)$$
In the continuous spectrum the eigenfunctions are $$
u(p,k,x)=\f{1}{\sqrt{2\pi}}\left (e^{ikx}-\f{p}{p+i|k|}e^{i|kx|}\right),\ -\infty <k< \infty, 
\eqno(3)$$
with energies $\hbar^2k^2/2m$ while the bound state energy is $W=-\hbar\omega_0=-\hbar^2p^2/2m$.
Functions $u(p,k,x),\ u_b(p,x)$ are normalized to $\delta(k-k')$ and to unity respectfully.
Parameter $m$ represents the mass of the bound charged particle.

On the interval $0\leq t\leq T$ acts a perturbing potential which is described by adding
in Eq.(1) a time dependent term $V(x,t)$ 
$$V(x,t)=eEx\eta(t),\ \ {\rm or}\ \ V(x,t)=R\d(x)\eta(t),\ \ t\in [0,T], \eqno(4)$$
where $\eta_{max}=1$ and parameters $E$, $R$ are responsible for the amplitude of perturbation.
Thus we have to solve the time-dependent Schr{\"o}edinger equation$$
i\hbar\f{\p\psi(x,t)}{\p t}=H_0\psi(x,t)+V(x,t)\psi(x,t),\ \ t\geq 0.\eqno(5)$$
After expanding $\psi(x,t)$ in the complete set of functions $u$: $$
\psi(x,t)=\theta(t)u_b(p,x)e^{i(\hbar p^2/2m)t}+\int^\infty_{-\infty}{\Theta(k,t)u(p,k,x)
e^{-i(\hbar k^2/2m)t}dk},\ \ t\geq 0,\eqno(6)$$
the survival of the bound state at time $t\leq T$ can be evaluated by $|\theta(t)|^2$, if we 
assume the system to be initially in its bound state$$
\theta(0)=1,\ \ \Theta(k,0)=0.\eqno(7)$$

It is more convenient [1] to proceed in dimensionless units ($\hbar=2m=g/2=1$) and rewrite$$
u_b(x)=e^{-|x|},\ \ W_b=-1,\ \ u(k,x)=\f{1}{\sqrt{2\pi}}\left(e^{ikx}-\f{e^{i|kx|}}{1+i|k|}\right),
\eqno(8)$$ 
where $W_b$ is the rescaled energy of the bound state. The energies of states $u(k,x)$ are 
$W(k)=k^2$ with multiplicity two for $k\neq 0$. These functions are normalized to $\delta (k-k')$, 
while the bound state $u_b(x)$ - to $1$.
\vskip0.3cm
\noindent
{\bf 2. DIPOLE FIELD PULSE PERTURBATION}
\vskip0.2cm
Beginning at $t=0$ a perturbing potential $Ex\eta(t)$ is applied to the atom and 
it stops at $t=T$. Here parameter $E$ represents the electric field of the perturbation whose 
frequency is $\omega$.

For solving Eq.(5) we use Eq.(6) and expand $\psi(x,t)$ on the interval $(0,T)$ in the complete set (8) of 
functions $u$:$$
\psi(x,t)=\theta(t) u_b(x)e^{it}+\int_{-\infty}^{\infty}{\Theta(k,t)u(k,x)e^{-ik^2t}dk}.\eqno(9)$$
Then using their orthonormality and assuming cutoff of the perturbation potential for large $|x|$
reduce dimensionless form of Eq.(5) to the following set$$
\dot{\theta}(t)=\f{4E\eta(t)}{\sqrt{2\pi}}\int_{-\infty}^{\infty}{\Theta(k,t)
\f{e^{-i(k^2+1)t}}{(k^2+1)^2}kdk},\eqno(10a)$$
$$\dot{\Theta}(k,t)=-\f{4E\theta(t)e^{i(k^2+1)t}\sin\omega t }{\sqrt{2\pi}(k^2+1)^2}k.\eqno(10b)$$
The evolution of $\Theta(k,t)$ is determined in Eq.(10b) by $\theta(t)$ only. 
Using Eq.(7), then integrating Eq.(10b) in time, and substituting the result into Eq.(10a) we 
obtain a single equation which describes the evolution of $\theta(t)$ $$
\dot{\theta}(t)=-\f{(4E)^2}{2\pi}\eta(t)\int_0^t{\theta(t')\eta(t')dt'}
\int_{-\infty}^{\infty}{\f{e^{-i(k^2+1)(t-t')}}{(k^2+1)^4}k^2dk},\ 0<t<T.\eqno(11)$$
 
The ionization probability of the bound state is$$
P(t)=1-|\theta(t)|^2,\ {\rm for}\ t\leq T,\eqno(12)$$
and it becomes constant $P(T)$ for all later times $t\geq T$. The internal integral over $k$ 
in Eq.(11) can be expressed in terms of Fresnel's integrals $$
S(t-t')=\int_{-\infty}^{\infty}{\f{e^{-i(k^2+1)u}}{(k^2+1)^4}k^2dk}=\f{3-4u^2}{48}A+iu
\f{3+4u^2-4iu}{24}B,$$
\vskip-0.2cm
where  $$\eqno(13)$$
\vskip-0.4cm
$$A=\pi [1-\Phi(\sqrt{iu})],\ \ B=e^{-iu}\sqrt{\pi/iu}-A,\ \ u=t-t'\geq 0.$$

\vskip0.3cm\noindent
{\bf Numeric realization of ionizing by dipole pulses}
\vskip0.3cm
We solve Eq.(11) for $\theta(t)$ by using the following technique, which is implied by a known method 
for the Volterra equations. The integral in time is approximated by the summation over discrete equal
subintervals defined by equidistant points on the interval $[0,t]$: $0,\delta, 2\delta,...,N\delta
=t$. As it is clear from Eqs.(7) and (11) $\theta(0)=1,\ \dot\theta(0)=0$, then we have 
$\theta(\d)\app \theta(0)+\delta\dot\theta(0)=1$. In this way $\theta(2\delta)=\theta(\delta)+
\delta\dot\theta(\delta)$ where $\dot\theta(\delta)$ is evaluated by the integral (11) without 
involving the end point at $t=2\delta$. And so on by consequent computations of $\theta(j\delta)$ 
via the terms with numbers $0,1,2,...,j-1$. This technique realizes an approximate quadrature of 
Eq.(11) and its precision becomes better when subintervals $\delta$ get shorter. This approximation 
replaces Eq.(11) with the sum$$
\dot\theta(n\delta)=-\f{(4E)^2}{2\pi}\delta\eta(n\delta)\sum_{j=0}^n{\theta(j\delta)
\eta(j\delta)S((n-j)\delta)},\eqno(14)$$
which allows to find approximately $\theta((n+1)\delta)=\theta(n\delta)+\delta\dot{\theta}
(n\delta)$. We study in this work sometimes quite long pulses, they require for
an acceptable precision long computing time as $\delta$ should be very small. This can be helped 
by improving the transition from $\theta(n\delta)$ to $\theta((n+1)\delta)$ by involving two 
derivatives of $\theta$:$$
\theta((n+1)\delta)=\theta(n\delta)+\delta\dot{\theta}(n\delta)+\f{\delta^2}{2}\ddot{\theta}
(n\delta).\eqno(15)$$
Differentiating Eq.(11) we have$$
\ddot{\theta}(t)=-\f{(4E)^2}{2\pi}\Big[\dot\eta(t)\int_0^t{\theta(t')\eta(t')dt'}
\int_{-\infty}^{\infty}{\f{e^{-i(k^2+1)(t-t')}}{(k^2+1)^4}k^2dk}$$

$$+\f{5\pi}{128}\theta(t)\eta^2(t)-i\eta(t)\int_0^t\theta(t')\sin\omega t'dt'
\int_{-\infty}^{\infty}{\f{e^{-i(k^2+1)(t-t')}}{(k^2+1)^3}k^2dk}\Big].$$
By expressing the last integral over $k$ in terms of Fresnel functions as$$ 
V(u)=\int_{-\infty}^{\infty}{\f{e^{-iu(k^2+1)}}{(k^2+1)^3}k^2dk}=\pi[1-\Phi(\sqrt{iu})]\f{1-4iu
+4u^2}{8}+e^{-iu}\sqrt{\pi iu}\f{1+2iu}{4},\eqno(16)$$
Eq.(15) can be rewritten in the following form
\begin{equation} \begin{split}\nonumber
&\theta((n+1)\d)=\theta(n\d)+\f{8\d^2E^2}{\pi}\Big\{\f{i\d}{2}\eta(n\d)\sum_{j=1}^n{\theta(j\d)\eta(j\d)V((n-j)\d)}\\
&\hskip11.5cm (17)\\
&-\f{5\pi}{256}\theta(n\d)\eta^2(n\d)-
\Big[\eta(n\d)+\f{\d}{2}\f{d\eta}{dt}(n\d)\Big]\sum_{j=1}^n{\theta(j\d)\eta(j\d)S((n-j)\d)}\Big\},
\end{split}\end{equation} 
\noindent
convenient for numeric computation of the ionization probability. 
\vskip0.2cm\noindent
{\bf A. Sin-wave pulses $\sin(\omega t)$}
\vskip0.3cm
The ionization of our atom by the sin-wave electric field excitation, which models the 
laser pulses, means that $\eta(t)=\sin(\omega t)$. For illustration everywhere in this work 
the 'short' harmonic pulses will have only five cycles, $N=5$, this can be easily realized now by 
experimental techniques [7-10]. For $\omega<1$ even $5$ cycles of oscillations last a relatively
long time, sometimes $T>100$ and the use of $\ddot\theta(t)$ is necessary. The results are 
exhibited in Fig.1, where time $t$ is measured in numbers of cycles of perturbation. 

\vskip0.5cm\hskip2cm
\epsfig{file=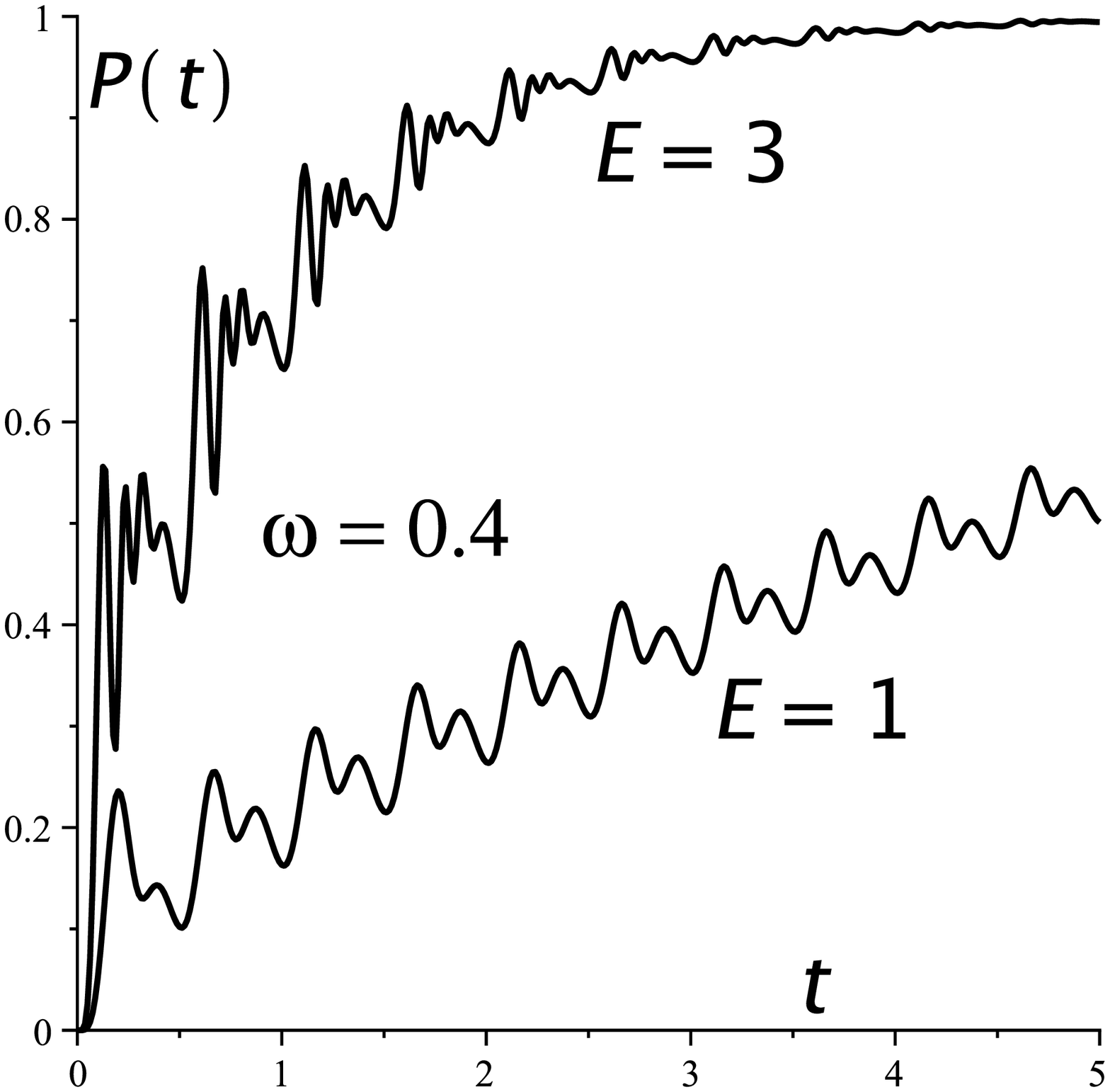, width=6cm, height=6cm} 

\vskip0.3cm
\centerline{\small FIG.1. Ionization probability $P(t)$ caused by harmonic pulses}
\vskip0.3cm\noindent
The plot structure shows that external frequency $\omega$ is doubled in the process, 
which is suggested in some measure by Eq.(11). When the electric field $E=3$ the complete
ionization occurs practically after the third cycle of the pulse. 

All computations are done using Maple on the intervals $T=5\f{2\pi}{\omega}$ with various $\omega$
and $E$, therefore $T=78.54, 104.7, 157.1$ for $\omega =0.4, 0.3, 0.2$ respectively. An acceptable
precision requires $\d\leq 0.04$, i.e. the sums in Eq.(17) will have up to $4000$ terms and
the computations is time consuming as the sum is of recursive nature.

\vskip0.5cm\hskip-0.6cm
\epsfig{file=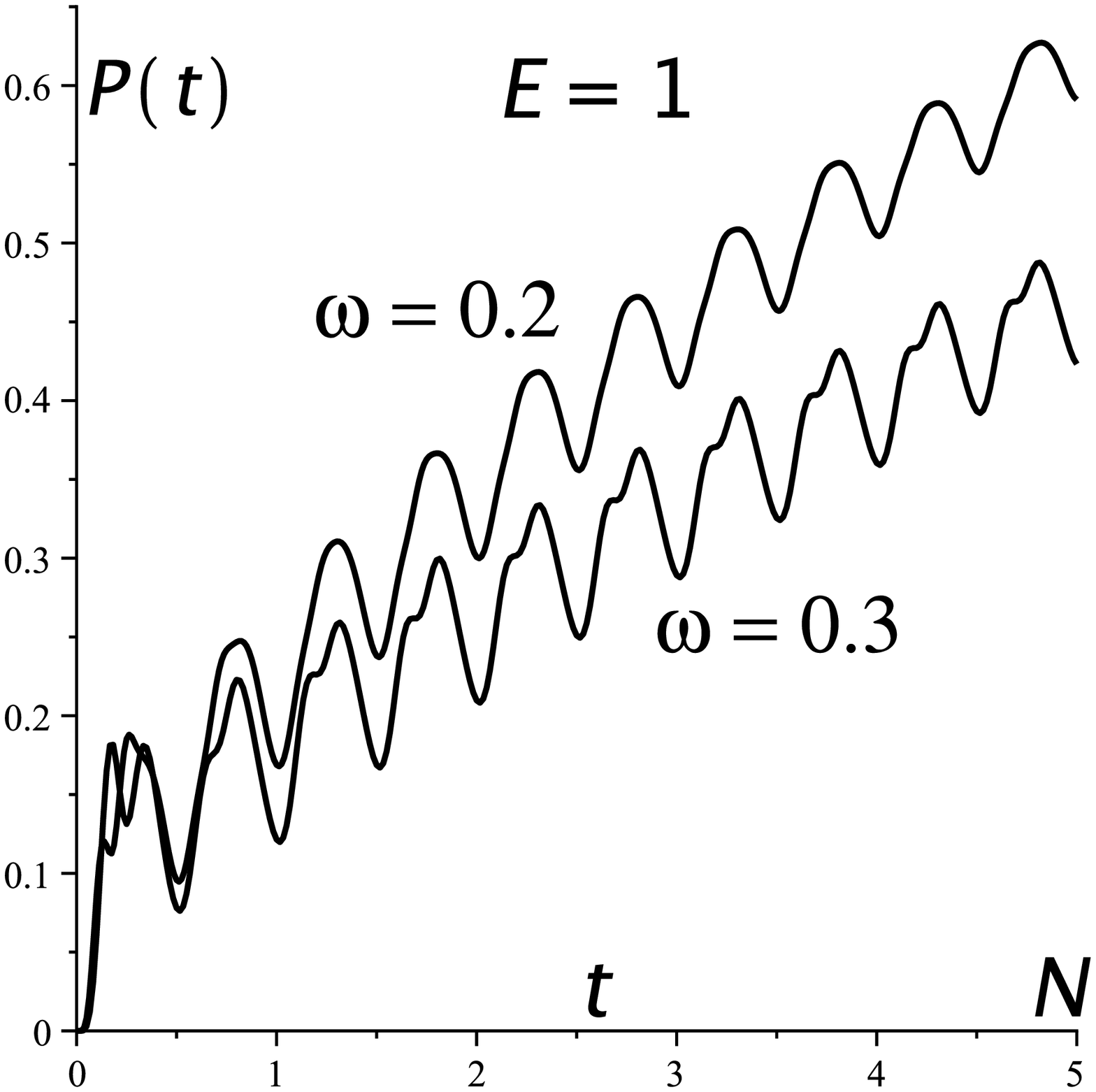, width=6cm, height=6cm} 

\vskip-6.3cm\hskip5.5cm
\epsfig{file=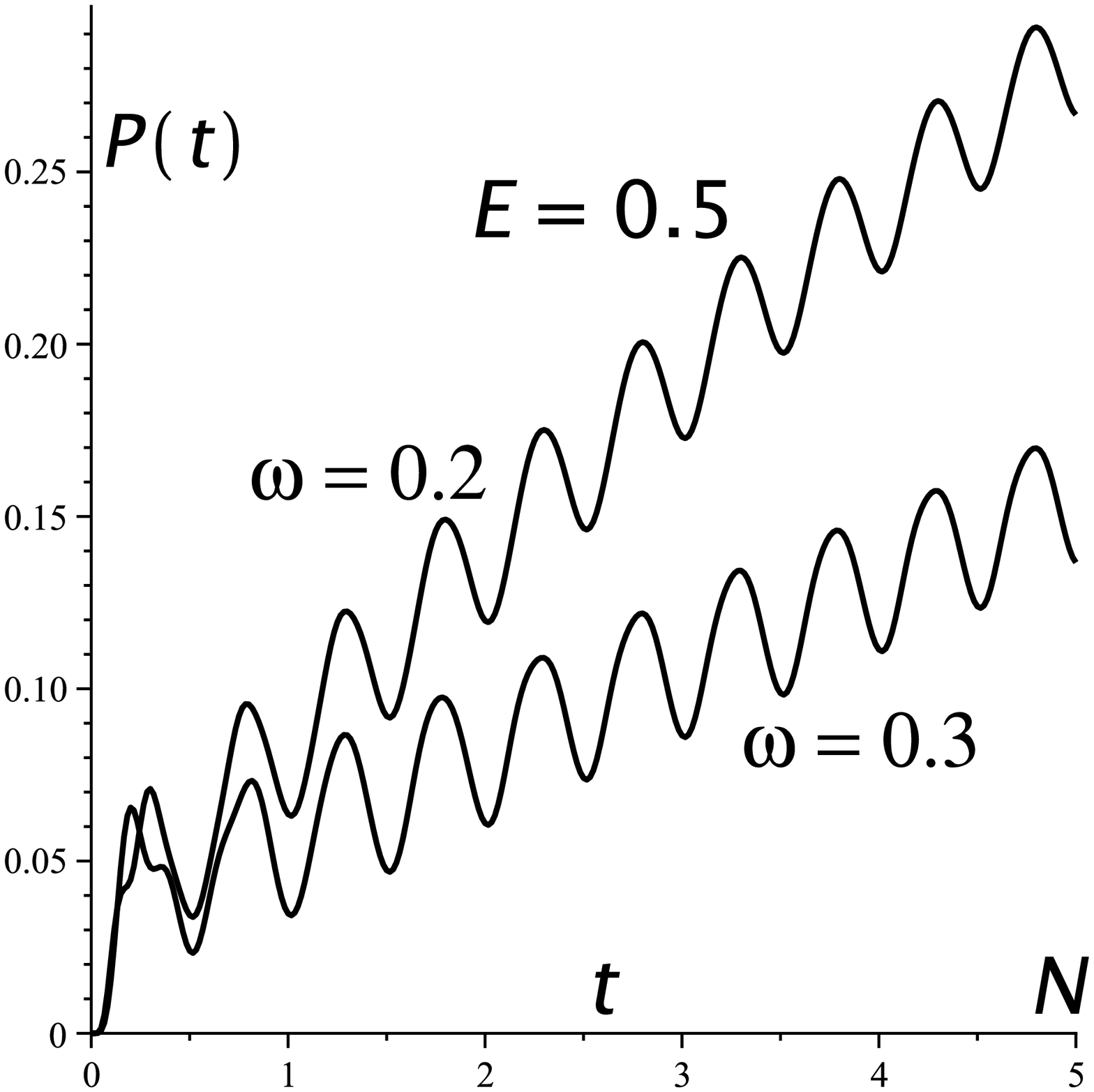, width=6cm, height=6cm} 

\vskip0.cm
\hskip1.5cm{\small FIG.2a. $E=1$}\hskip4cm{\small FIG.2b. $E=0.5$} 
\vskip0.2cm
{\small Fig.2. Ionization probability by five cycled harmonic pulses with $\omega =0.2$ and $0.3$}
\vskip0.3cm\noindent
Fig.2 confirms our observation that when $\omega$ is far from the resonance the total duration of
the perturbation pulse can be more important than its frequency, experiments [7-9] confirm this 
effect in real systems. The same behavior is more visible in Fig.2b, where the electric field is 
smaller $E=0.5$. 

For a very qualitative comparison with experimental results we note that in the case of Cesium
atoms, whose external orbitals have radii about 2 Angstroms, where the electric field is
about $30\ GV/m$. When we rescale $|W_{Cs}|=3.89\ eV$ to $|W_b|=1$ this would correspond in Eq.(14)
to $E\sim 8$ in the dimensionless units, therefore our $E=0.5$ means roughly about $2\ GV/m$, which
is a very high electric field in laser pulses, but reachable for present techniques. The
case with the field strength $E=0.2$ (in physical units this corresponds $E\sim 0.8\ GV/m$ for Cs 
or $\sim 1.6\ GV/m$ for Tungsten with $|W|=7.86\ eV$) was computed too: $P(T)\app 0.053$ for 
$\omega=0.2$ and $0.027$ for $\omega=0.3$ while the shape of curves $P(t)$ is similar to plots in 
Fig.2. These results are only for orientation and clearly should be considered as qualitative due
to the limitations of our model. It looks that for smaller $E$ the value of $P(T)$ becomes 
proportional to $\sim E^2$ in agreement with Eq.(17).
\vskip0.2cm\noindent
{\bf B. Pulsed dipole forcing}
\vskip0.3cm
Here we consider electric bursts acting on the model atom: rectangular $\eta(t)=1$ and bell-shaped
$\eta(t)=4(t/T-t^2/T^2)$ on the interval $0\leq t\leq T$. Though the bell-shaped pulse has the
same amplitude as the rectangular one, but its ionization efficiency is much lower in Fig.3 because
its total energy is smaller and more importantly it does not have high frequency harmonics. The
ionization probability $P(t)$ is an oscillating function but there is an important difference
between plots in Figs.3a and 3b. While the ionization probability by a rectangular pulse of 
duration $t_1$ is given by $P(t_1)$ in Fig3a, the ionization evolution for bell-shaped pulses  
is presented by Fig.3b, but only for the pulse of length $10$, i.e. if $T=t_1$ function $P(t_1)$ 
in Fig.3b does not represent the final probability and the corresponding computation should be 
performed namely for $T=t_1$ because $\d=\d(T)$. Our calculation with $E=0.2$ and the bell-shaped 
pulse is not plotted in Fig.3b because the curve $P(t)$ runs very low: its maximum $P(4)\sim 0.013$ 
and $P(10)$ is less than $0.002$.

\vskip0.5cm\hskip-0.6cm
\epsfig{file=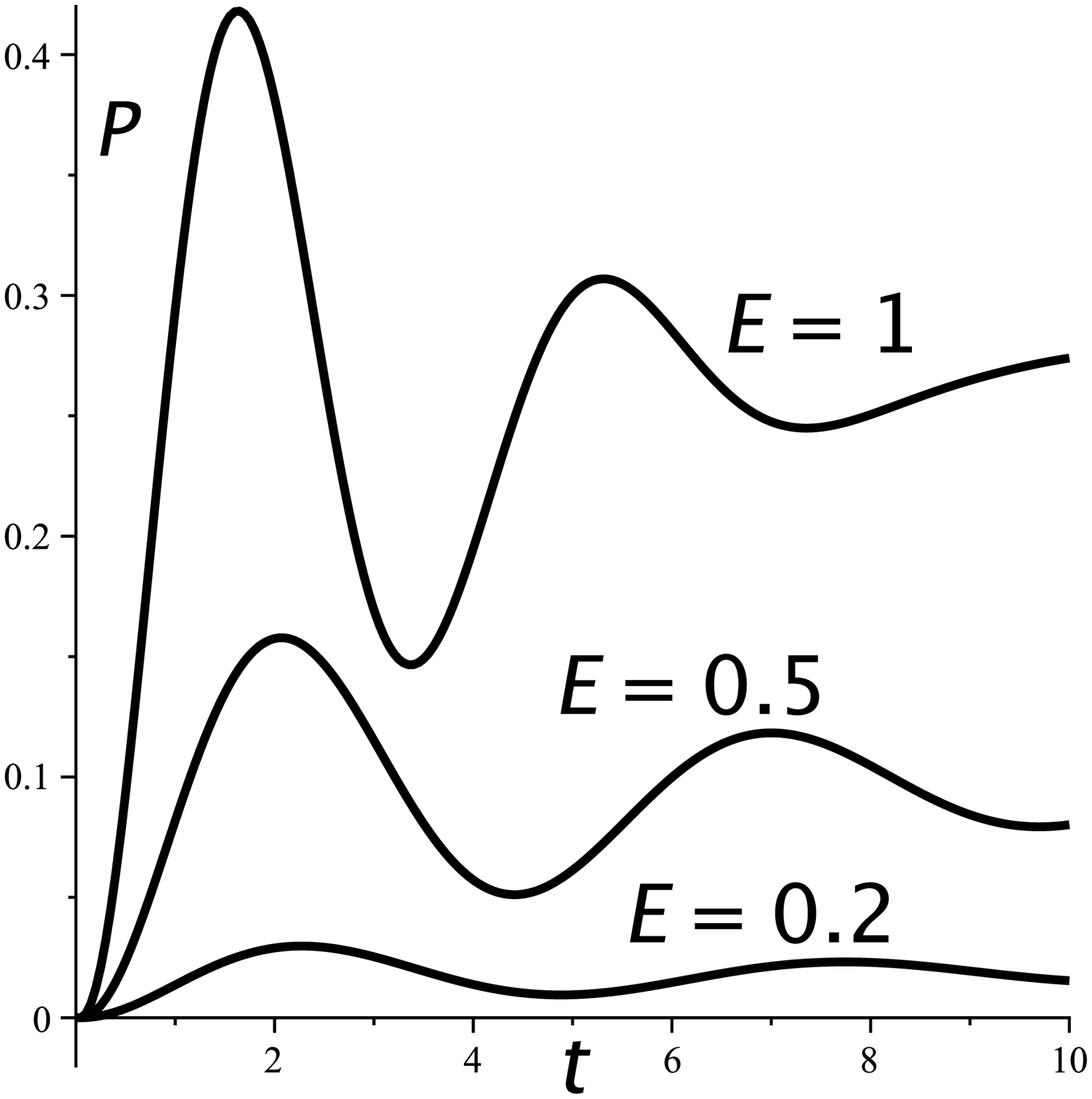, width=6cm, height=6cm} 

\vskip-6.2cm\hskip5.5cm
\epsfig{file=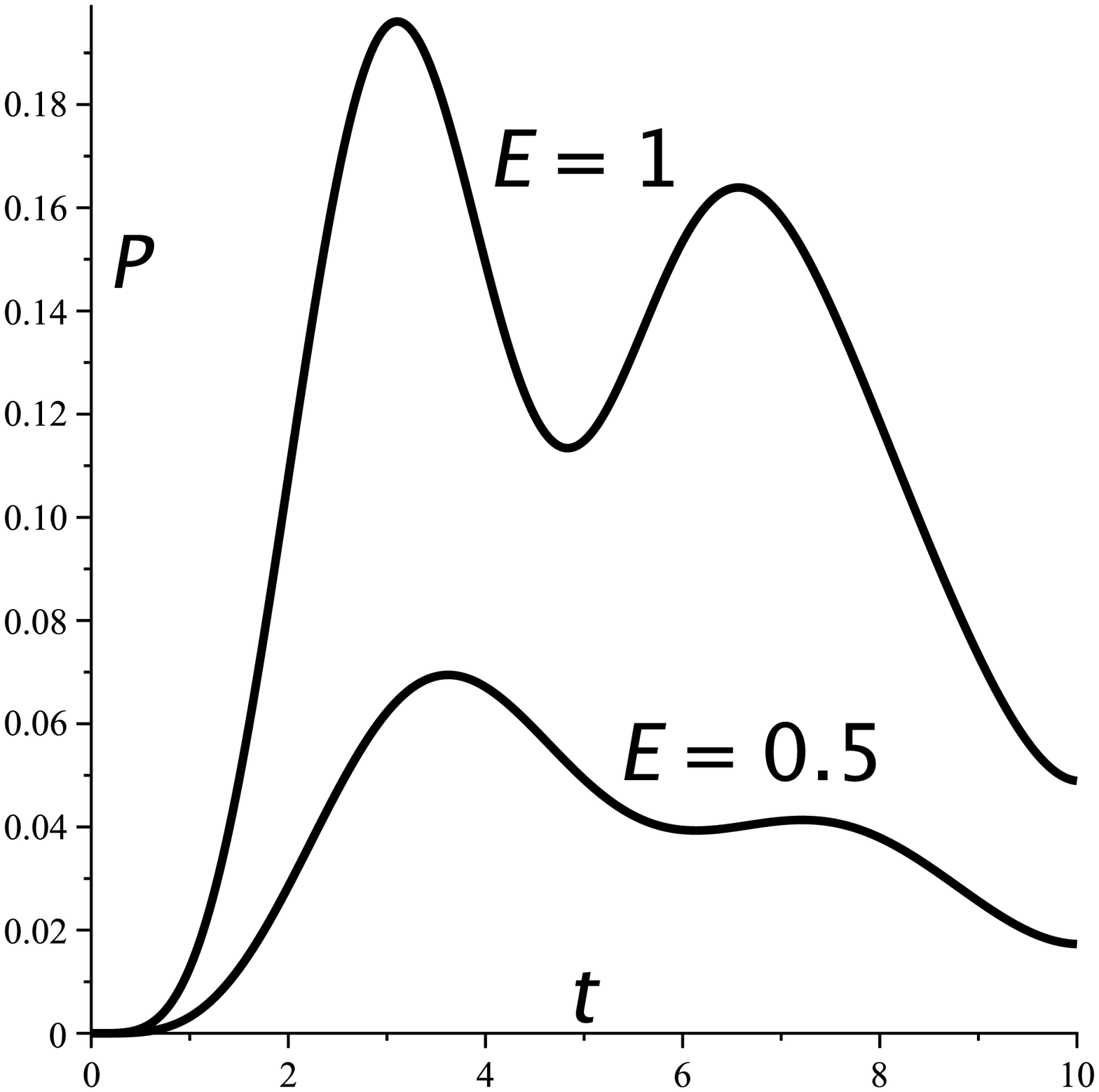, width=6cm, height=6cm} 

\vskip0.cm
\hskip0.7cm{\small FIG.3a. Rectangular pulse}\hskip2cm{\small FIG.3b. Bell-shaped pulse} 
\vskip0.1cm
{\small Fig.3. Ionization probability caused by rectangular and bell-shaped electric bursts}
\vskip0.3cm\noindent

Our approach is modified below for studying the ionization caused by the parametric modulations of 
the binding potential. Though this is hardly achievable in practice but, as we already mentioned, 
it exhibits some illuminating features of the ionization by short pulses which are quite universal. 
\vskip0.3cm
\noindent
{\bf 3. PULSED MODULATION OF BINDING POTENTIAL.}
\vskip0.2cm

Here we consider short pulse bursts of the potential strength which has to be studied by a 
different computations technique. Eq.(5) now has the following form$$
i\f{\p\psi(x,t)}{\p t}=H_0\psi(x,t)+R\d(x)\eta(t)\psi(x,t),\ \ \eta(t)=0\ {\rm when}\ 
t\notin [0,T],\ \ t\geq 0,\eqno(18)$$
with the initial conditions given by Eq.(7). Using the expansion (6) in terms of the stationary 
eigen-functions and the methods developed in [1] yields a simple equation for the function 
$\theta(t)$ 
$$\theta(t)=1+2i\int_0^t{Y(t')dt'},\eqno(19)$$
where $Y(t)$ is defined by the following integral equation$$
Y(t)=R\eta(t) \left\{1+\int_0^t{[2i+M(t-t')]Y(t')dt'}\right\}.\eqno(20)$$
The function $M(s)$ in (20), see [1], is$$
M(s)=\f{2i}{\pi}\int_0^\infty{\f{u^2e^{-is(1+u^2)}}{1+u^2}du}=-i+\sqrt{\f{i}{\pi s}}e^{-is} +
i\Phi(\sqrt{is}).\eqno(21)$$
It behaves as $\sqrt{i/\pi s}-i$ when $s\to 0$ and is proportional to $s^{-3/2}e^{-is}$ 
when $s\to\infty$.
\vskip0.3cm
\noindent
{\bf A. Case of rectangular pulse $\eta(t)=1$.}
\vskip0.2cm

For approximate evaluation of the function $\theta(t)$ in this set up we have to solve 
numerically the integral equation (20) that can be done if there is an effective way of 
computing $M(s)$ for not very large values of $s$ as our pulses are not long. Using [11] equation 
for the integral in Eq.(21) can be written in terms of gamma functions$$
\int_0^\infty{\f{\sqrt{x}e^{-isx}}{1+x}dx}=e^{is}\Gamma\left(\f{3}{2}\right)\Gamma\left(-\f{1}
{2},is\right),$$
and then one can apply the power series expansion for the confluent hyper-geometric function and
present $M(s)$ in a rapidly convergent form$$
M(s)= -i-\sqrt{\f{i}{s\pi}}\sum_{n=0}{\f{(-is)^{n}}{(2n-1)\ n!}}. \eqno(22)$$
For a given pulse duration $T$ the upper limit in the sum (22) can be only slightly larger 
than $eT$ to give a good precision for $M(s)$.

Our next step is express this function by a power series using the behavior of $Y(t)$ near zero,$$
Y(t)=\sum_{m=0}{c_mt^{m/2}},\ \ 0<t<T,\eqno(23)$$
and substitute Eqs.(22) and (23) into the integral equation (20). The result reads$$
R^{-1}\sum_{m=0}{c_mt^{m/2}}=1+i\sum_{m=0}{c_m\f{t^{1+m/2}}{1+m/2}}+\sum_{k,n=0}   
{B\left(\f{k}{2}+1,n+\f{1}{2}\right)c_ka_n t^{n+\f{k+1}{2}}},$$
where $B(x,y)=\Gamma(x)\Gamma(y)/\Gamma(x+y)$ is the $\beta$-function [11]. Thus we obtain  
equations for finding coefficients $c_*$:$$
c_0=R,\ \ \ c_1=B\left(1,\f{1}{2}\right)Ra_0c_0=2R^2\sqrt{\f{i}{\pi}},$$
\vskip-0.8cm
$$\eqno(24)$$
\vskip-0.8cm
$$c_m=\f{2iR}{m}c_{m-2}+R\sum_{n=0}^{\lfloor\f{m-1}{2}\rfloor}{B\left(\f{m+1}{2}-n,n+\f{1}{2}
\right)a_n c_{m-2n-1}},\ \ m\geq 2.$$
The symbol $\lfloor x\rfloor$ as usual means the integer part of $x$, and coefficients 
$c_m$ in (24) can be calculated consequently as they are expressed in terms of $c_*$ evaluated 
earlier. Clearly we will keep in (22) and (23) only finite numbers of terms.

By combining Eqs.(19), (23), and (12) we find the ionization probability as a function of  
the pulse duration $T$ $$
P(T)=1-\Big|1+4i\sum_{m=0}{\f{c_m}{m+2}T^{1+m/2}}\Big|^2.\eqno(25)$$
\noindent
The results of computation by Eqs.(22-25), where we keep about 80 terms in sums, are presented
in Figs.4 which show this probability for different values of pulse amplitude $R$ and $T\leq 10$. 
Left Fig.4 shows also that a very short attractive pulse with $R=-0.5,\ T\app 0.9$ is more 
effective for ionization than longer pulses of the same amplitude. 
\vskip0.5cm\hskip-0.6cm
\epsfig{file=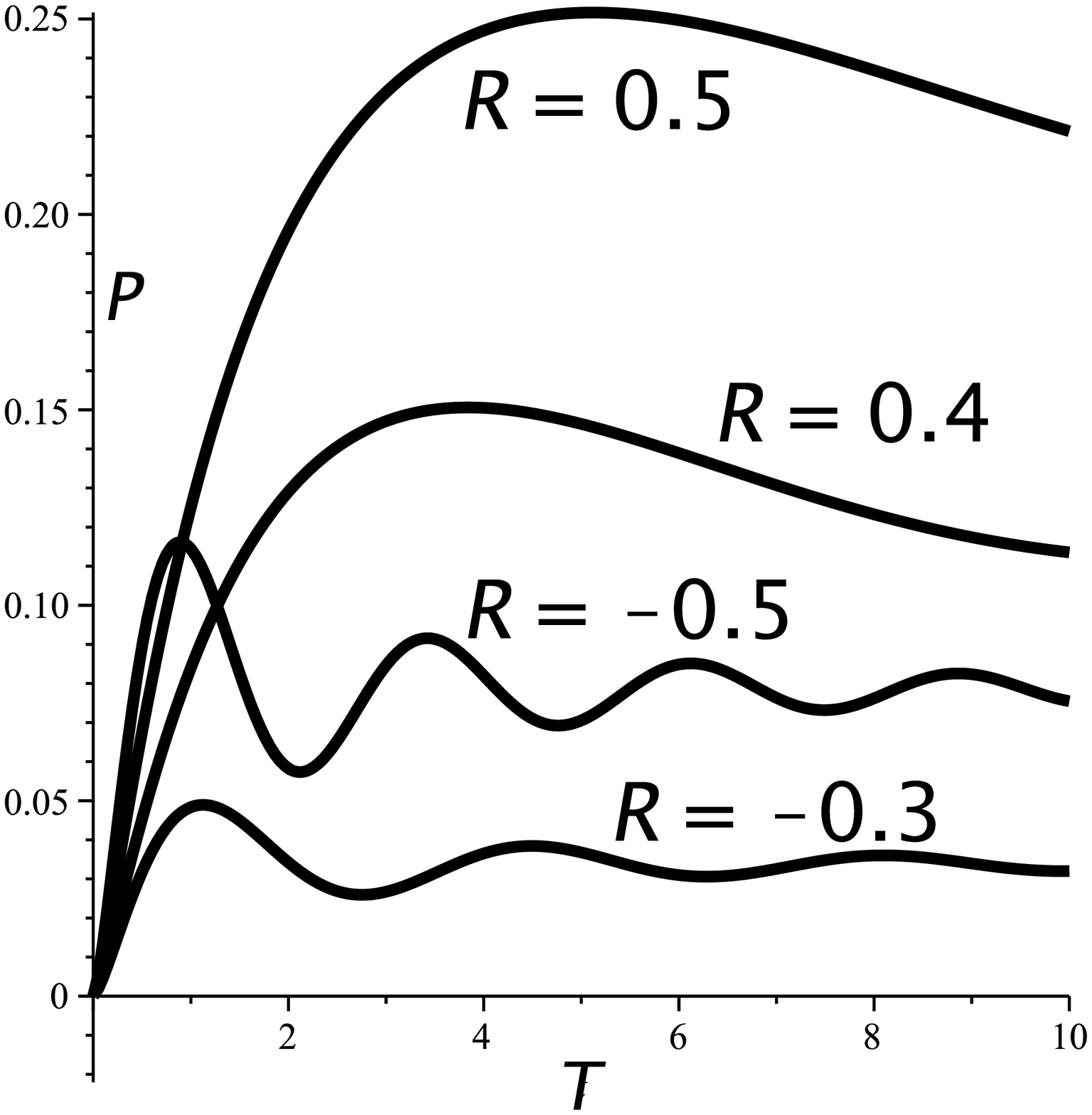, width=6cm, height=6cm} 

\vskip-6.0cm\hskip5.5cm
\epsfig{file=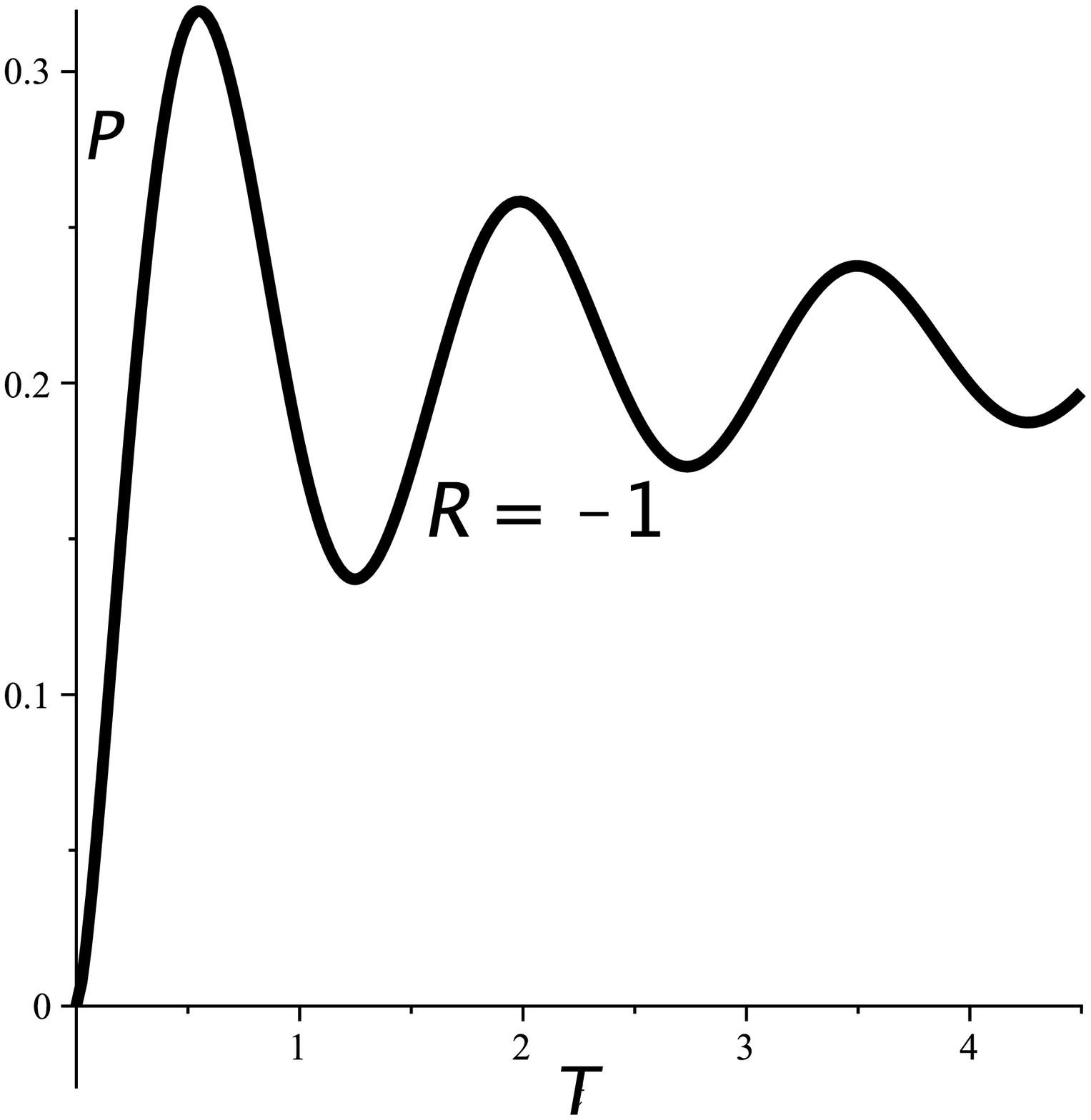, width=6cm, height=6cm} 

\vskip0.cm
\centerline{\small FIG.4 Plots of ionization probability $P(T)$ for rectangular pulses} 
\vskip0.2cm
In agreement with the common sense when the external pulse decreases the binding energy, $R>0$ in 
(4), the ionization rate is larger. There is always an intermediate pulse duration corresponding 
to maximum of ionization and for longer positive pulses the ionization rate approaches to a 
constant. Right Fig.4 exhibits the ionization probability caused by the attractive positive pulse 
whose amplitude is one half of the binding energy and duration $T\leq 4.5$. One can see the 
resonances and quite a high probability $0.32$ when $T=0.55$ because the pulse fronts have infinite 
slopes and thus introduce very high harmonics though the total pulse energy is limited. Note also 
that $R$ should be compared with $g=2$ in our units, $|R|=1$ is the largest considered here.

\bigskip\noindent
{\bf B. Bell-shaped pulse $\eta(t)=4(t/T-t^2/T^2)$}

\medskip
On the interval $0<t<T$ we consider a pulse, symmetric about $t=T/2$, whose maximum amplitude
is $R$, see Eq.(4). In this case the same approach as before converges slower and one needs more
terms in Eqs.(22) and (23), we used up to 400-500 of them.

A straightforward analysis of Eq.(20) shows that the power series expansion for $Y(t)$ has the 
following form$$ 
Y(t)=\sum_{m=0}{c_mt^{1+m/2}},\ \ 0<t<T.\eqno(26)$$
By substituting (26) into Eq.(20) and using Eq.(22) we solve Eq.(26) and find the coefficients
which define $Y(t)$  $$
\f{Tc_0}{4R}=1,\ c_1=0,\ \f{Tc_2}{4R}=-\f{1}{T},\ \f{Tc_3}{4R}=a_0c_0B\left(\f{1}{2},2\right),\ 
\f{Tc_4}{4R}=i\f{c_0}{2}+a_0c_1B\left(\f{1}{2},\f{5}{2}\right),$$ 
$$\f{Tc_5}{4R}=i\f{2c_1}{5}+a_0c_2B\left(\f{1}{2},3\right)+a_1c_0B\left(\f{3}{2},2\right)-
\f{1}{T}a_0c_0B\left(\f{1}{2},2\right),$$ 
\vskip-0.3cm
$$\eqno(27)$$
\vskip-0.8cm
$$\f{Tc_m}{4R}=\f{2ic_{m-4}}{m}-\f{2ic_{m-6}}{T(m-2)}+a_0c_{m-3}B\left(\f{1}{2},\f{m+1}{2}
\right)$$
$$+\sum_{n=0}^{\lfloor\f{m-5}{2}\rfloor}{c_{m-2n-5}\left(\f{2n+1}{m}a_{n+1}-\f{a_n}{T}\right)
B\left(n+\f{1}{2},\f{m-1}{2}-n\right)},\ \ m\geq 6.$$
One can see again that as in Eqs.(24) each coefficient $c_*$ is defined via corresponding ones
found earlier.

\vskip0.5cm\noindent
\epsfig{file=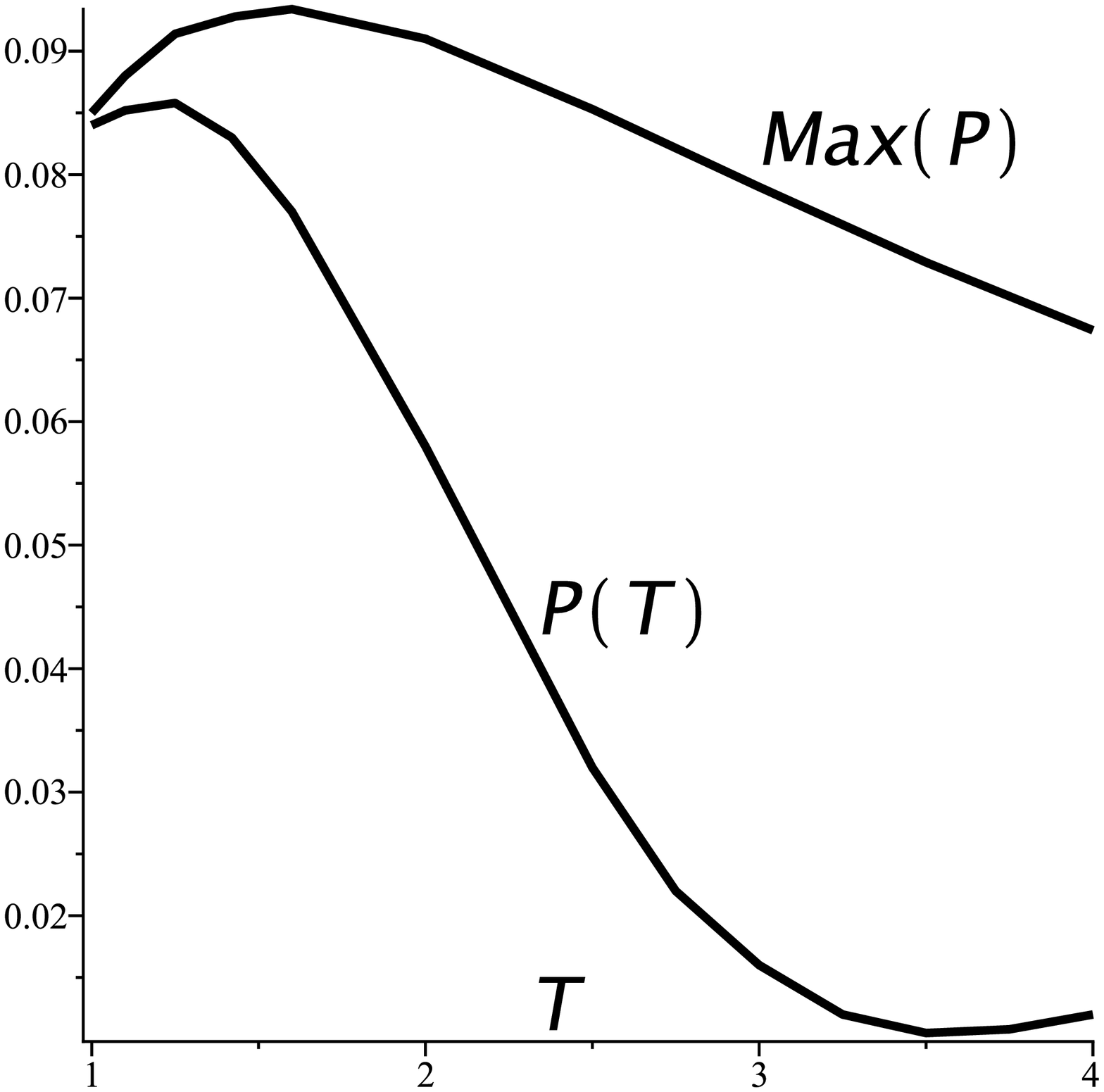, width=6cm, height=6cm}
\vskip-5.8cm\hskip5.5cm
\epsfig{file=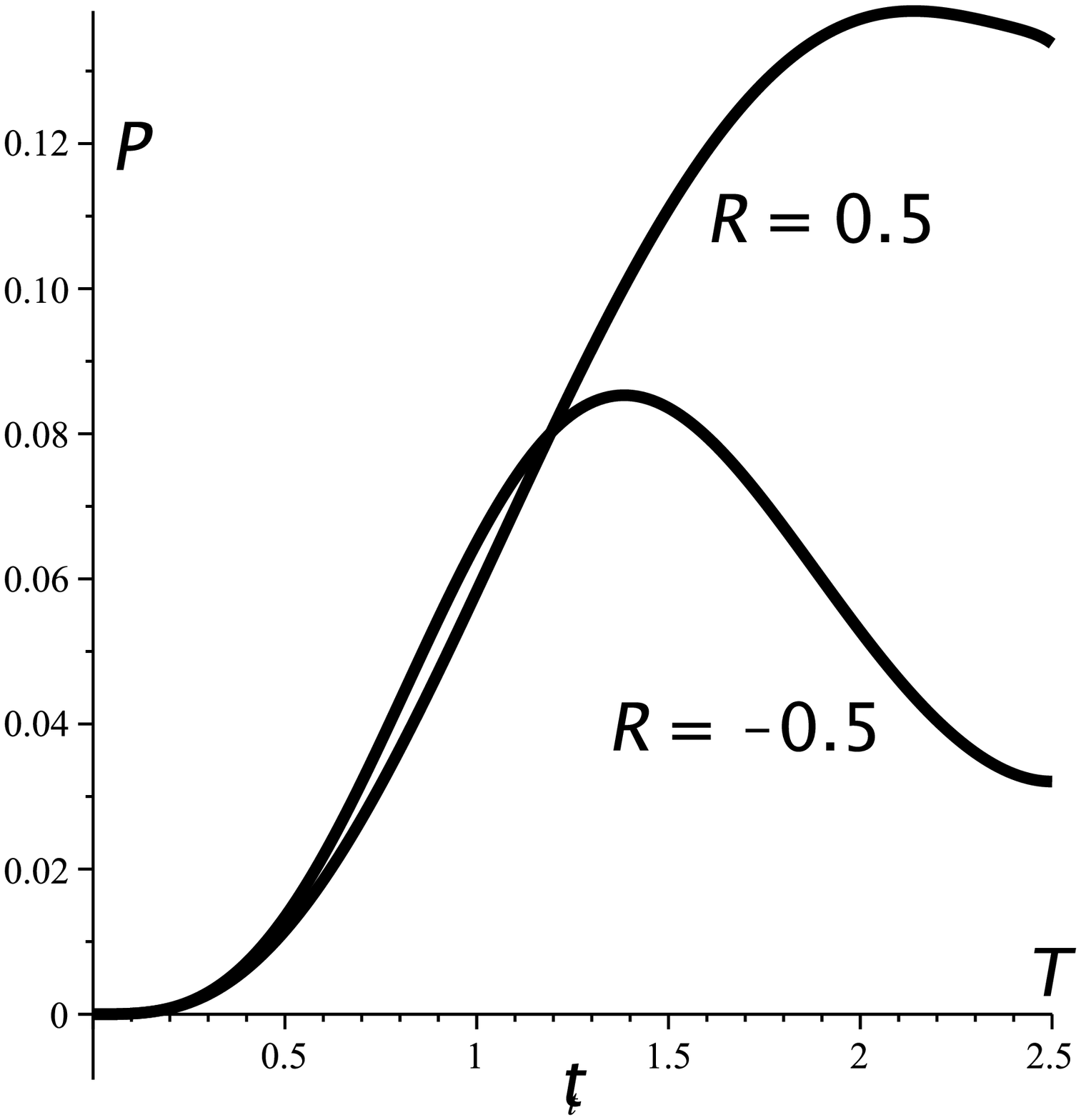, width=6cm, height=6cm} 
\vskip-0.2cm\hskip2cm
{\small FIG.5a}\hskip5cm {\small FIG.5b}

{Ionization probability by bell-shaped}\hskip1.2cm {Time evolution of ionization

~~{ pulses of different duration $T$} \hskip1.9cm {when $T=2.5$ and $R=\pm 0.5$} 

\vskip0.3cm\noindent
Note that in the case of a rectangular pulse after solving Eqs.(24) the solution can by found 
by using Eqs.(19) and (20) for any pulse duration $t=T$ or in Fig.5a. But for the present 
bell-shaped pulse the meaning of plots in Fig.5b is different, see Part 2B, and also the 
ionization probability is found using Eqs.(26), (27) and (25) in the slightly modified form$$
P(T)=1-|\theta(T)|^2=1-\Big|1+4i\sum_{m=0}{\f{c_m}{m+4}T^{2+m/2}}\Big|^2,\eqno(28)$$
where all $c_k,\ k=0,1,...$ depend on $T$.

$P(T)$ in Fig.5a is the ionization probability at the end of the corresponding pulse with $R=-0.5$ 
and length $T$, but the plot for maximum $Max(P)$ which occurs at an intermediate time $0<t<T$ 
and for longer pulses it is much larger than $P(T)$. Fig.5b shows $P(t)$ behavior in the case 
$T=2.5$ when the pulse amplitudes are $R=\pm 0.5$. Positive pulses are 
more effective as before: $P(T)=0.13$ is close to $Max(P)$ when $T=2.5$ while $P(T)$ and $Max(P)$ 
are $0.139$ and $0.147$ respectively for $T=3$. This probably means that change of sign of 
the external force plays an important role and might suggest that more realistic perturbation 
(a harmonic one with some envelope) can be more efficient. One can see that the system can be 
ionized before the end of perturbation pulse, i.e. the value of $Max(P)$ is interesting.

\bigskip\noindent
{\bf C. Short sin-wave pulses $\eta(t)=\sin(\omega t)$}

\medskip
Our pulse exists on the interval $0\leq t\leq T$ and as in Part 2 its length has always an integer 
number $N=5$ of cycles, $T=2\pi N/\omega$. We assume in computations that the frequency of 
principal harmonic of $Y(t)$ is $\omega$ and choose an integer $K$ of harmonics sufficient for 
modeling $Y(t)$. Then using the Galerkin [12] method for solving Eq.(20) the function $Y(t)$ is 
approximated by the following sum$$
Y(t)=\sum_{k=-K}^K{a_k f_k(t)},\ \ {\rm where}\ \ f_k(t)=e^{i\omega kt}.\eqno(29)$$
The solution method requires the discrepancy of using the approximation (29) in Eq.(20) be 
orthogonal to all functions $f_k$. This procedure creates the linear algebraic system for 
coefficients $a_k$ $$
\sum_{k=-K}^K {a_k\int_0^T{dt \bar f_m(t)\Bigg\{f_k(t)-R\eta(t)\int_0^t{[2i+M(t')]f_k(t-t')
dt'}\Bigg\}}}=$$
\vskip-0.6cm
$$\eqno(30)$$
\vskip-0.6cm
$$R\int_0^T{\eta(t)\bar f_m(t)dt},\ \ -K\leq m\leq K.$$

By substituting Eq.(29) into (30) this system can be rewritten in the standard form$$
\sum_{k=-K}^K{C_{k,m}a_k}=B_m,\eqno(31)$$
where all coefficients with $k\neq m\pm 1$ defined by the following relations:$$
C_{k,m}=A_{k-m}+\f{R}{2\omega}\left(\f{M_{-k}-M_{1-m}}{k-m+1}-
\f{M_{-k}-M_{-1-m}}{k-m-1}\right).\eqno(32)$$
For $k=m-1$ and $k=m+1$ we have respectively$$
C_{m-1,m}=\f{R}{4\omega}[(1+2i\omega T)M_{1-m}-M_{-1-m}-2i\omega\overline M_{1-m}],$$
\vskip-0.6cm
$$\eqno(33)$$
\vskip-0.6cm            
$$C_{m+1,m}=\f{R}{4\omega}[(1-2i\omega T)M_{-1-m}-M_{1-m}+2i\omega\overline M_{-1-m}].$$

Other terms in Eqs.(31-33) are
$$A_{k-m}=
\begin{dcases} 
0, & k\neq m, \\ 
T, & k=m, 
\end{dcases}\hskip0.7cm B_m=\f{R}{2i}[A_{1-m}-A_{-1-m}],$$
\vskip-0.6cm
$$\eqno(34)$$
\vskip-0.8cm
$$M_n=\int_0^T{[2i+M(t)]e^{in\omega t}dt},\hskip0.6cm \overline M_n=
\int_0^T{[2i+M(t)]te^{in\omega t}dt}.$$

Our computation by Eqs.(31-34) for the cases when $T=5$, $R=0.2,\ 0.4$, and the pulse 
consisting of $5$ cycles (therefore $\omega = 2\pi N/T\app 6.3>\omega_0$) are shown in 
Fig.6: 
\vskip0.5cm\hskip2cm
\epsfig{file=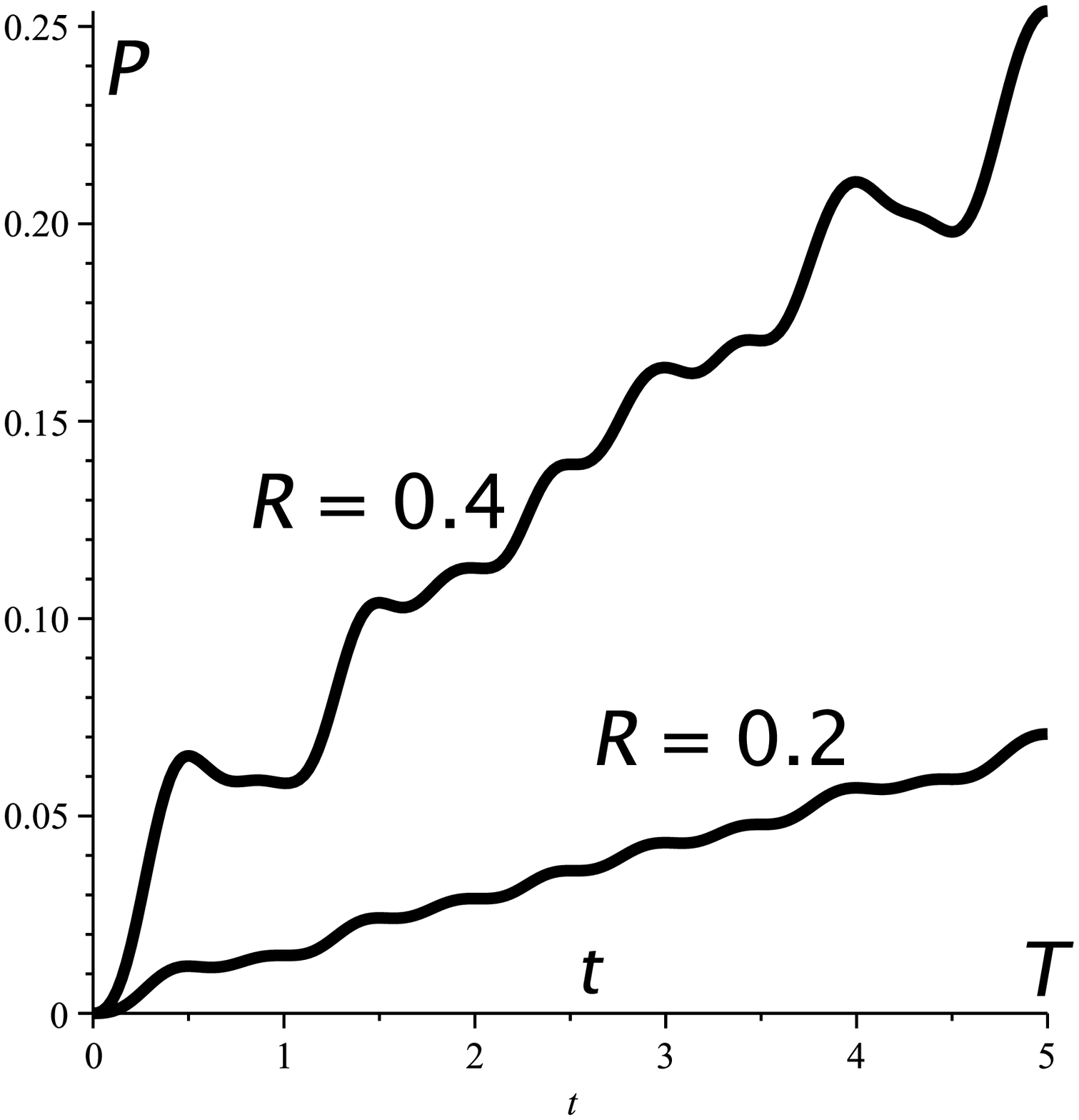, width=6cm, height=6cm} 

\vskip0.3cm
\centerline{\small FIG.6. Ionization probability $P(T)$ caused by sin-wave pulse}
\vskip0.5cm\noindent
One can see that with pulse amplitude $0.4$ the ionization is quite effective. We used 
only five harmonics, $K=5$, in computations but checked the precision by taking $K=10$ 
which produced the curves $P(t)$ almost identical to ones in Fig.4. The roughly linear 
time dependence of the ionization probability is caused by the fact that $\omega$ is 
significantly larger than $\omega_0$, i.e. the energy of ionizing photons in Fig.4 exceeds 
the binding energy of the model atom and thus we observe the first order effect. In 
reality this process might be compared with the soft X-ray ionization. 

For solving Eq.(20) this method of approximation $Y(t)$ by Eq.(29) and the routine defined
by Eqs.(31-34) becomes inefficient when $\omega$ is much smaller because the interval $T$ 
is too long in this case and oscillations of $M(s)$ are not well modeled by harmonics of 
$\eta(t)$. We use the special properties of the Volterra Eq.(20) to model function $Y(t)$ 
by a set of its discrete points. When the ionization is provided by at least two photons
and $\omega=0.6$, 5 cycles of sin wave make $T>60$, and thus one needs about $N=10^3$ points 
for a decent approximation of $Y(t)$. Denoting temporarily $F(s)=2i+M(s)$ the integral 
equation (20) is replaced by the following one, where we keep the integral term in the 
interval of divergent behavior of $M(s)$ $$
R^{-1}Y(t_n)=\eta(t_n)+\eta(t_n)\left[\Delta \sum_{m=1}^{n-1}{F(t_n-t_m)Y(t_m)}+
\int_0^{\Delta}{F(s)Y(t_n-s)ds}\right].\eqno(35)$$
Here $t_n=n\Delta$, $n$ runs from $1$ to $N$ and $\Delta=T/N$. The integral in Eq.(35) is
approximated using Eq.(22) for $M(s)$ when $s$ is small and only the linear term of 
$Y(t)$-dependence on the interval $(t_{n-1},t_{n})$:$$
F(s)\app i+\sqrt{\f{i}{\pi s}}\left[1+is+\f{s^2}{6}-i\f{s^{3}}{30}\right],
\ Y(t_n-s)\app Y(t_{n})+\f{Y(t_{n-1})-Y(t_{n})}{\Delta}s.\eqno(36)$$
Near small $s$ we keep more terms in singular $F(s)$ than in $Y(t_n-s)$ because $Y(t)$
has a regular behavior there. By evaluating the integral in Eq.(35) we come to the recurrent
equation for computing the set $Y(t_n)$ consequently starting from $n=2$. $$
Y(t_n)=\f{R\eta(t_n)}{1-B-i\Delta/2}\left[1+\Delta \sum_{m=1}^{n-1}{F(t_n-t_m)Y(t_m)}+(A+i\Delta/2)
Y(t_{n-1})\right],\eqno(37)$$
$$
A=\sqrt{\f{i\Delta}{\pi}}\left(\f{2}{3}+\f{2i\Delta}{5}+\f{\Delta^2}{21}-
\f{i\Delta^3}{135}\right),\ B=\sqrt{\f{i\Delta}{\pi}}\left(\f{4}{3}+
\f{4i\Delta}{15}+\f{2\Delta^2}{105}-\f{2i\Delta^3}{945}\right).$$
Here we will neglect terms of the order $\Delta^6$ and higher.

It is clear that $Y(0)=0$ and for approximating $Y(t_1)$ by a polynomial with the same precision we 
substitute $Y(\Delta-s)$ into Eq.(20) and using Eqs.(36,37) obtain the following relation$$
\sum_{k=0}^{10}{c_k\Delta^{1+k/2}}=R\left(\omega\Delta-\f{\omega^3\Delta^3}{3!}+\f{\omega^5\Delta^5}
{5!}\right)\left\{1+\sum_{n=0}^{6}c_n\left[\f{i\Delta^{2+n/2}}{2+n/2}+\sqrt\f{i}{\pi}H_n(\Delta)
\right]\right\},$$
where $H_n(\Delta)=$ $$\eqno(38)$$ $$
\Delta^{\f{3+n}{2}}\left[B\left(\f{4+n}{2},\f{1}{2}\right)+i\Delta B\left(\f{4+n}{2},
\f{3}{2}\right)+\f{\Delta^2}{6}B\left(\f{4+n}{2},\f{5}{2}\right)-\f{i\Delta^3}{30}
B\left(\f{4+n}{2},\f{7}{2}\right)\right].$$
These equations are sufficient to evaluate $c_k,\ k=0,...,10$ and find $Y(t_1)=Y(\Delta)$. We 
applied Eqs.(35-38) to compute the ionization by the five cycle pulses of lower than in Fig.6
frequency $\omega=0.6$ and amplitudes $R=0.2,\ R=0.4$.
\vskip0.cm\hskip2cm
\epsfig{file=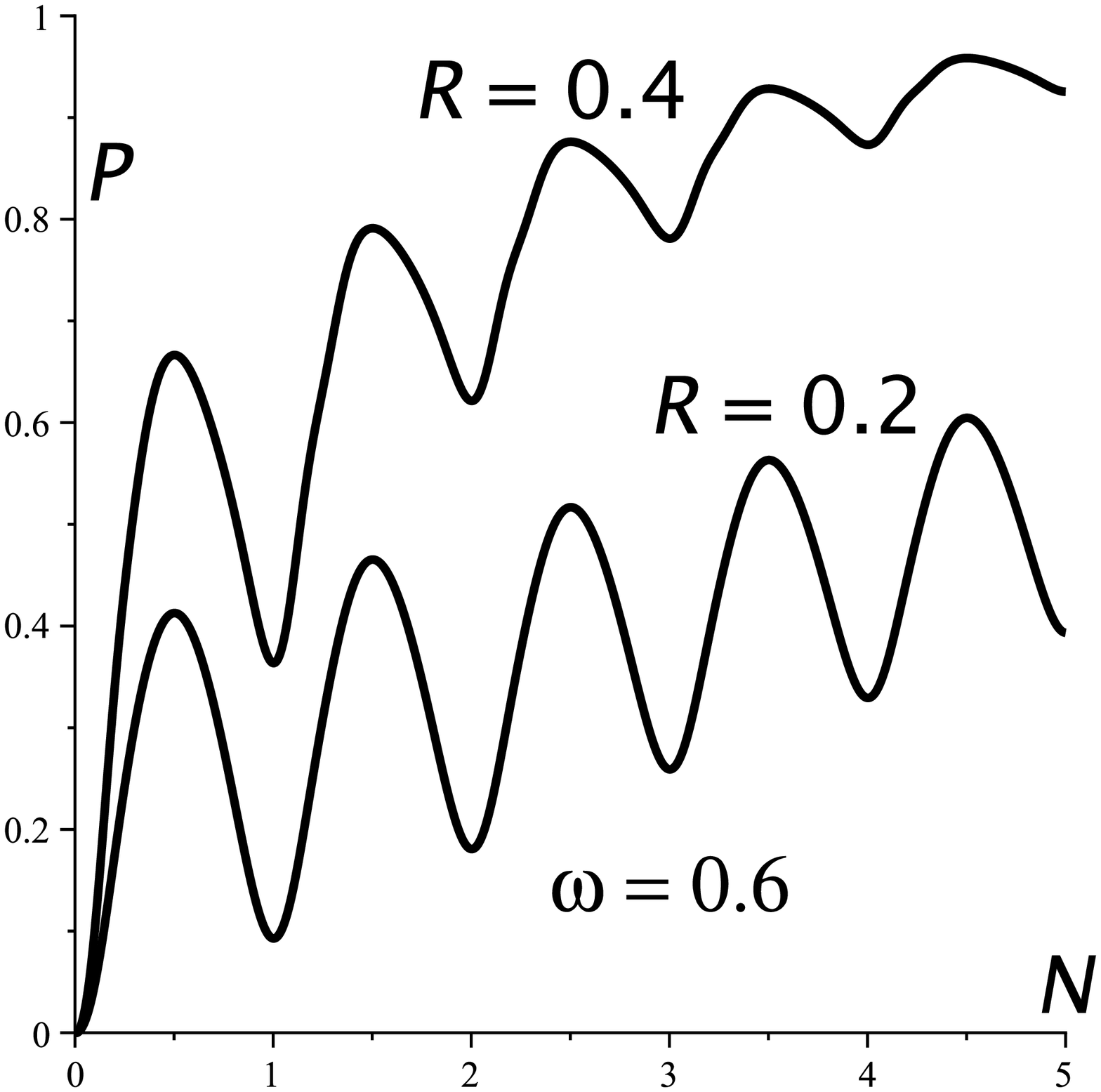, width=6cm, height=6cm} 

\vskip0.3cm
\centerline{\small FIG.7. Ionization initiated by sin-wave pulse with $\omega=0.6$}
\vskip0.5cm\noindent
The dynamics of the process is shown in Fig.7. One can see that the ionization probability is 
higher than in the case of $\omega\app 6$ in Fig.6. This result is probably caused by a much 
longer (more than in order) action of the perturbation while its average force (determined by $R$)
roughly the same. The sin-wave pulses appeared to be quite efficient for ionization and $P(T)$ 
seems to be proportional to $R^2$ for smaller $R$ like for the laser pulse perturbation. The time 
of perturbation in Fig.5 is measured in the number of harmonic cycles, the pulse ends when $N=5$ 
and $T\app 52.4$. 

In the dimensionless units when the atomic ionization energy is $1$ our results can be tentatively
mapped onto real system again, say Tungsten W and Cesium Cs with binding energies $7.86\ eV$ and 
$3.89\ eV$ respectively. It is easy to see that $\omega=0.6$ would correspond the wavelength about 
$260\ nm$ for W and $530\ nm$ for Cs in our cases. In experiments often are used $\sim 4-10$ fsec 
laser pulses of $\lambda =800-830\ nm$, [7-10], and having in mind a qualitative application of 
our theory we perform a somewhat less precise computation for smaller $\omega=0.4$ and $0.2$ 
respectively when the pulse time $T$ is longer.
\vskip0.5cm\hskip2cm
\epsfig{file=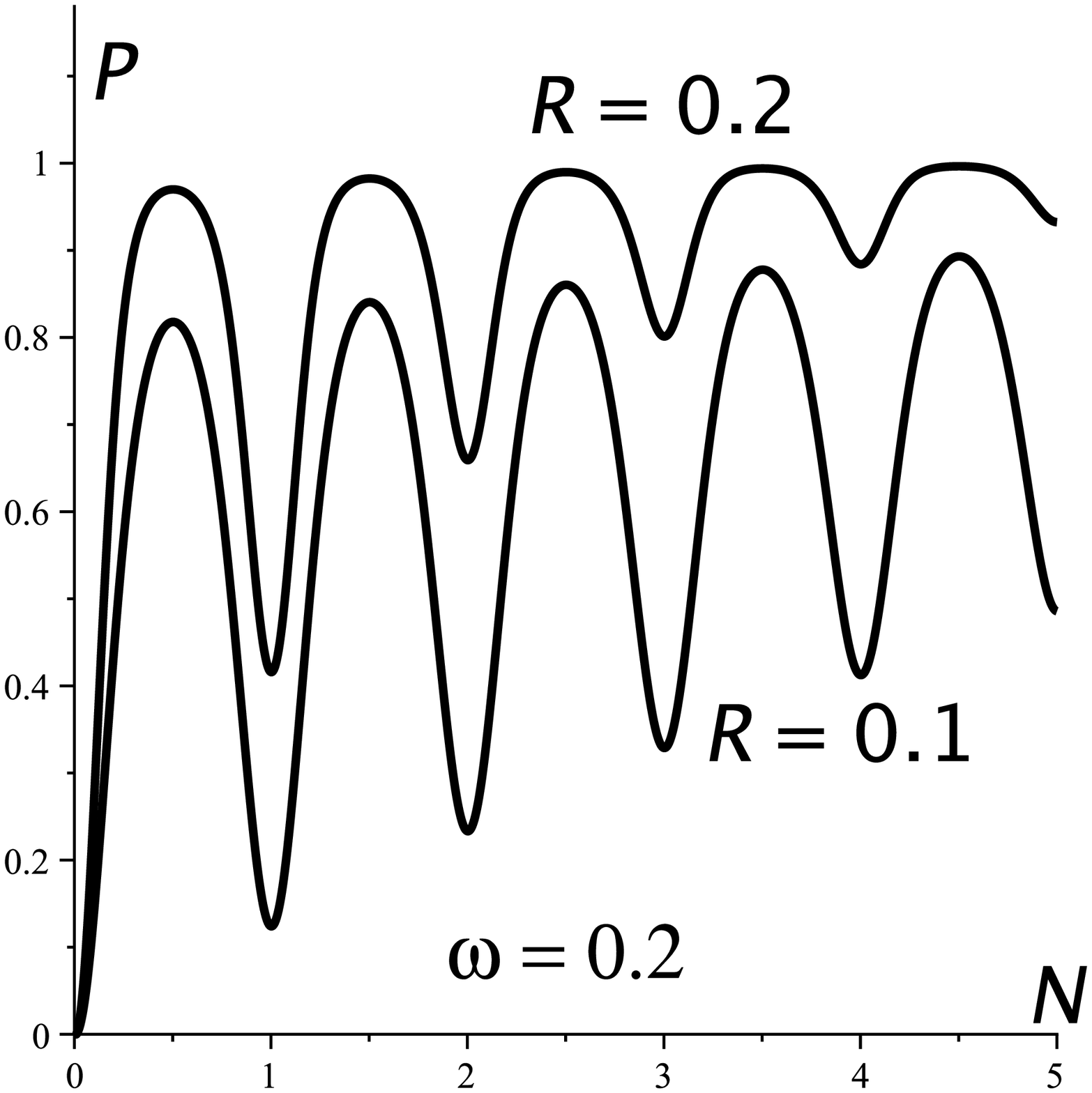, width=6cm, height=6cm} 

\vskip0.1cm
\centerline{\small FIG.8. Ionization by short pulse harmonic waves of $\omega=0.2$}
\vskip0.5cm\noindent
The results presented in Figs.8 and 9 confirm the importance of total pulse duration.
As before our pulses have only $5$ cycles, the time of their action is $T=157$ dimensionless units
when $\omega=0.2$ in Fig.6 and twice shorter when $\omega=0.4$ in Fig.9. Fig.8 roughly imitates 
the Cesium ionization when $R=0.1$ and $0.2$ while $\lambda\sim 800-830\ nm$. The amplitude $R=0.2$
is sufficient for almost complete ionization even at the pulse beginning. 

Fig.9 gives a hint of the Tungsten ionization (keeping in mind the same $\lambda\sim 800\ nm$).

\vskip0.5cm\hskip2cm
\epsfig{file=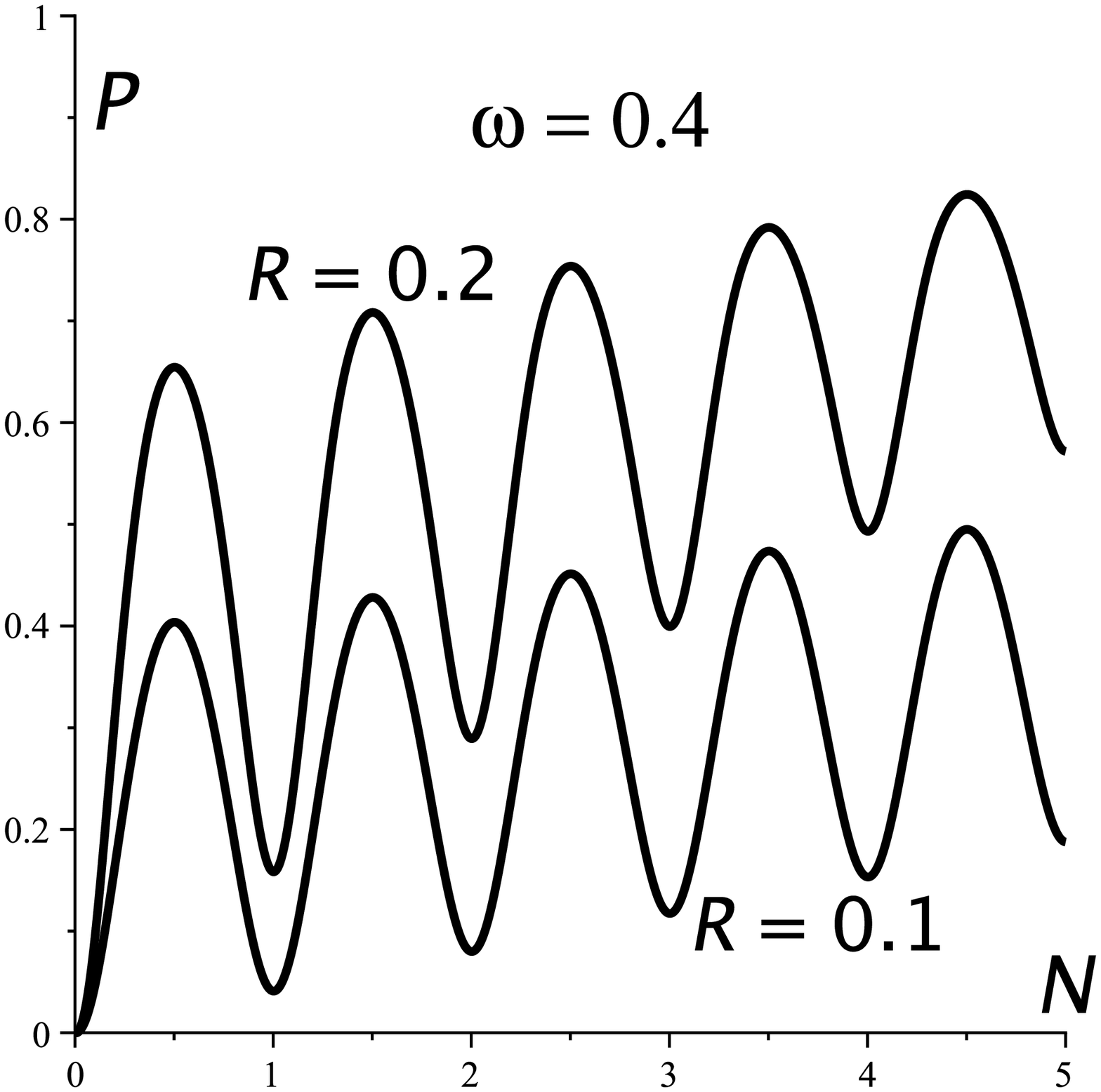, width=6cm, height=6cm} 

\vskip0.1cm
\centerline{\small FIG.9. Short pulse ionization when atomic binding energy is larger}
\vskip0.5cm\noindent
The ionization level is lower than in case of $Cs$ though the wave length is twice shorter. This 
clearly agrees with greater binding energy in $W$ atoms.

As the parametric perturbation acts directly on the binding energy it is more efficient in 
Figs.7-9 than the more realistic perturbation by the external harmonic electric field in Part 2.
These results show that the ionization of our model atom in some measure describes qualitative 
behavior of real systems.

\bigskip\noindent
{\bf 4. SUMMARY}
\vskip0.3cm

The atomic ionization by short pulses of external forces is studied on a simple one-dimensional 
model which allows to construct an exact theory of the process and realize its conclusions by
several methods of numerical computations. This creates a basis for comparison with approximate 
solutions of more realistic models, simulations, and experiments. Our main results include the
observation that for external frequencies, much lower than the resonance ones, the total duration 
of the pulse is more important for effective ionization than its frequency. When the ionization 
level is substantially far from the complete one it is increasing approximately linearly in time 
and has resonances as a function of pulse duration. For ionization caused by the dipole electric
field the frequency of these resonances is twice larger than the frequency of external forcing.

\end{document}